\documentclass[superscriptaddress,nobalancelastpage,10pt,pra,letter]{revtex4}

\usepackage{amsmath}
\usepackage{amsthm}
\usepackage{amssymb}
\usepackage{amsfonts}
\usepackage{graphicx}
\usepackage{wasysym}
\usepackage{mathrsfs}
\usepackage{yfonts}
\usepackage{bbold}
\usepackage{verbatim}
\usepackage{subfigure}
\usepackage{color}
\usepackage{soul} 
\usepackage{lipsum}
\usepackage{notes2bib}
\usepackage{colortbl}

\makeatletter
\newcommand*{\rom}[1]{\expandafter\@slowromancap\romannumeral #1@}
\makeatother

\begin{document}


\title{Resolution limit in quantum imaging with undetected photons using position correlations}

\author{Balakrishnan Viswanathan}
\affiliation{Department of Physics, Oklahoma State University, 145 Physical Sciences Bldg., Stillwater, Oklahoma 74078, USA}
\author{Gabriela Barreto Lemos} 
\affiliation{Instituto de F\'isica, Universidade Federal do Rio de Janeiro, Av. Athos da Silveira Ramos 149, Rio de Janeiro, CP: 68528, Brazil.}
\author{Mayukh Lahiri}
\email{mlahiri@okstate.edu} \affiliation{Department of Physics, Oklahoma State University, 145 Physical Sciences Bldg., Stillwater, Oklahoma 74078, USA}

\begin{abstract}
Quantum imaging with undetected photons (QIUP) is a unique method of image acquisition where the photons illuminating the object are not detected. This method relies on quantum interference and spatial correlations between the twin photons to form an image. Here we present a detailed study of the resolution limits of position correlation enabled QIUP. We establish a quantitative relation between the spatial resolution and the twin photon position correlation in the spontaneous parametric down-conversion process (SPDC). We also quantitatively establish the roles that the wavelength of the undetected illumination field and the wavelength of the detected field play in the resolution. Like ghost imaging and unlike conventional imaging, the resolution limit imposed by the spatial correlation between twin photons in QIUP cannot be further improved by conventional optical techniques.
\end{abstract}

\maketitle

\section{Introduction}
Quantum imaging uses the quantum states of light to image beyond the classical capabilities. In addition to overcoming the classical limits of sensitivity \cite{brida2010experimental,sabines2019twin,garces2020quantum,Casacio2021} and spatial resolution \cite{schwartz2013superresolution,classen2017superresolution,unternahrer2018super,tenne2019super}, one important achievement of quantum imaging is the discovery of fundamentally new imaging techniques such as interaction free imaging \cite{white1998interaction}, ghost imaging  \cite{klyshko1988effect,belinskii1994two,pittman1995optical,gatti2008quantumimaging,chan2009two,aspden2013epr, moreau2019imaging}, and, more recently, \emph{quantum imaging with undetected photons} (QIUP)  \cite{lemos_quantum_2014,lahiri2015theory,viswanathan2021position}. QIUP requires spatially correlated twin photons that are usually produced by spontaneous parametric down-conversion (SPDC) in nonlinear crystals. However, no coincidence measurement or post-selection is performed to acquire the image, which marks the distinctive feature of QIUP. In particular, the photons illuminating the object are not detected and the image is acquired solely by the detection of photons that never interacted with the object. Since the illumination and detection wavelengths can be very different, this  enables us to image at wavelengths for which any sort of detector is not accessible.
\par
Recently, QIUP has drawn a lot of attention. Notably, its application to mid-infrared microscopy \cite{kviatkovsky2020microscopy,paterova2020hyperspectral} has shown one way of circumventing the need for inefficient and expensive infrared detectors. However, there is only limited understanding of the resolution of QIUP. 
In fact, most experimental work published thus far uses the far field configuration, in which the imaging is enabled by the momentum correlation of the twin photons \cite{lahiri2017twin,hochrainer2017quantifying}. Besides, the understanding of resolution also exists only in this domain \cite{fuenzalida2020resolution}. In order to assess the full range of capabilities of QIUP, it is essential to have an understanding of the resolution beyond this domain. 
\par
Here we present a detailed analysis of the spatial resolution of QIUP in the near field configuration. In this case, the imaging is enabled by the position-correlation between the twin photons \cite{viswanathan2021position}. We derive a quantitative relationship between the resolution and the position correlation between the twin photons. We also establish a quantitative relationship between the resolution and wavelengths of the twin photons. We illustrate our results with numerical simulations and highlight the differences between the resolution limits in the near field (position correlation enabled) and far field (momentum correlation enabled) cases.
\par
This article is organized as follows. In Sec. \ref{sec:basic-theory}, we provide a brief description of the imaging scheme and recollect the basic theory involved. In Sec. \ref{sec:reso-method}, we present a detailed theoretical analysis along with the results. More specifically, we study the point spread function and the minimum resolvable distance. In this section, we also discuss the dependence of resolution on position correlation and wavelengths of the twin photons. In Sec. \ref{sec:discuss}, we compare the resolution of position correlation enabled and momentum correlation enabled QIUP schemes. We also point out the similarities between the resolution of QIUP and that of quantum ghost imaging. Finally, we summarize and conclude in Sec. \ref{sec:concl}.

\section{Basics of the theory of position correlation enabled QIUP}\label{sec:basic-theory}
\begin{figure}
	\centering 
	\includegraphics[width=1
	\linewidth]{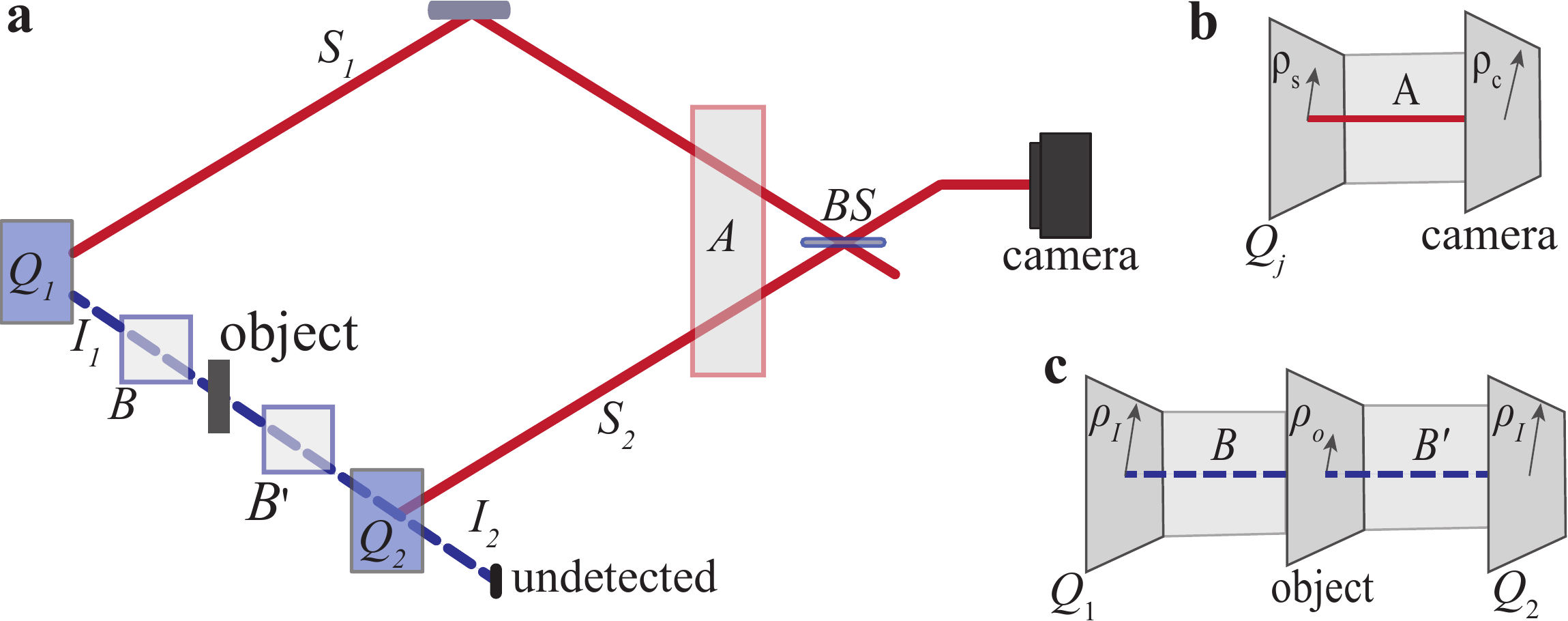}
	\caption{\textbf{a}, Illustration of the imaging scheme. Two identical twin photon sources, $Q_1$ and $Q_2$, can emit non-degenerate photon pairs (signal and idler) into beams $(S_{1},I_{1})$ and $(S_{2},I_{2})$. An imaging system, $B$ images the idler field at $Q_1$ onto the object, $O$, with a magnification $M_I$. Another imaging system $B'$ images the idler field at the object onto $Q_2$ with a magnification $1/M_I$. Idler beams $I_1$ and $I_2$ (dashed lines) are perfectly aligned and never detected. Signal beams $S_1$ and $S_{2}$ (solid lines) are superposed by a $50:50$ beam-splitter ($BS$) and projected onto a camera. An imaging system $A$ ensures that the signal field at the sources is imaged onto the camera with a magnification $M_S$. The image of $O$ is obtained from the single-photon interference patterns observed at the camera without any coincidence measurement or post-selection. \textbf{b}, A point $\boldsymbol{\rho}_{s}$ located on $Q_j~(j=1,2)$ is mapped onto a point $\boldsymbol{\rho}_{c}=M_s\boldsymbol{\rho}_{s}$ on the camera by the imaging system $A$. \textbf{c}, Due to the presence of the imaging systems $B$ and $B'$, a point $\boldsymbol{\rho}_{I}$ on $Q_1$ is mapped onto a point $\boldsymbol{\rho}_{o}=M_I\boldsymbol{\rho}_{I}$ on the object and then again at $\boldsymbol{\rho}_{I}$ on $Q_2$.}
	\label{fig:img-schm}
\end{figure}

A general schematic of QIUP in the near field configuration \footnote{This scenario must not be confused with conventional near-field imaging where evanescent fields play an integral role. We stress that evanescent fields play no role in our imaging scheme.} is given in Fig. \ref{fig:img-schm}. There are two identical sources, $Q_1$ and $Q_2$, each of which can produce a photon pair. The two photons belonging to a pair are called signal ($S$) and idler ($I$). Suppose that $Q_1$ emits the signal and the idler photons into beams $S_1$ and $I_1$, respectively (Fig. \ref{fig:img-schm}a). Likewise, $S_2$ and $I_2$ represent the beams into which the signal and the idler photons are emitted by $Q_2$. The two signal beams, $S_{1}$ and $S_{2}$, are superposed by a $50:50$ beamsplitter (BS) and one of the outputs of the BS is detected by a camera. The beam $I_1$ from source $Q_1$ illuminates the object, passes through source $Q_2$, and gets perfectly aligned with beam $I_2$. This alignment makes the which-way information unavailable and as a result the two signal beams, $S_{1}$ and $S_{2}$, interfere. The phenomenon is sometimes called \textit{induced coherence without induced emission}, as the effect of stimulated emission due to the alignment of idler beams is negligible \cite{zou1991induced,wang1991induced,wiseman2000induced,lahiri2019nonclassical}. A conceptual discussion of this phenomenon can be found in \cite{lahiri2020partially}.
\par
An object ($O$) placed on the path of the idler photon between $Q_1$ and $Q_2$ introduces the which-way information and consequently affects the interference pattern generated by the signal photons. This fact allows us to retrieve the information about the object from the interference pattern. We stress that no idler photon is ever detected and the image is constructed by detecting only the signal photons which never interact with the object. 
\par
Both the object and the camera are placed in the near field relative to $Q_1$ and $Q_2$ by the use of appropriate imaging systems. We choose a general setup that provides a complete understanding of the spatial resolution. An imaging system ($A$) with a magnification $M_S$ ensures that the signal field at the sources is imaged onto the camera (Fig. \ref{fig:img-schm}a,b). An imaging system, $B$, is placed between the source $Q_1$ and the object ($O$) in the beam $I_1$ such that the idler field at $Q_1$ is imaged onto the object with a magnification $M_I$. Another imaging system, $B'$, images the idler field at the object onto source $Q_2$ with a magnification $1/M_I$ (i.e., demagnified by an equal amount). These two imaging systems also ensure that $Q_2$ lies in the image plane of $Q_1$. The image is obtained from the interference pattern observed on the camera. In this case, the imaging is enabled by the position correlation between the signal and idler photons \cite{viswanathan2021position}.
\par
We now briefly recollect the theory of position correlation enabled quantum imaging with undetected photons. A detailed description can be found in Ref. \cite{viswanathan2021position}.
\par
The quantum state generated by each source individually can be represented by
\begin{equation} \label{state-spdc}
|\psi\rangle = \int d\textbf{q}_{s}  \ d\textbf{q}_{I} \ C(\textbf{q}_{s}, \textbf{q}_{I}) |\textbf{q}_{s}\rangle_{s} |\textbf{q}_{I}\rangle_{I},
\end{equation}
where $|\textbf{q}_{s}\rangle_{s}$ and $|\textbf{q}_{I}\rangle_{I}$ denote a signal photon with transverse momentum $\hbar \textbf{q}_{s}$ and an idler photon with transverse momentum $\hbar \textbf{q}_{I}$, respectively. The complex quantity $C(\textbf{q}_{S}, \textbf{q}_{I})$ ensures that $|\psi\rangle$ is normalized.
\par
The joint probability density of detecting the signal and the idler photons at positions (transverse coordinates) $\boldsymbol{\rho}_{s}$ and $\boldsymbol{\rho}_{I}$, respectively, on the source plane is given by \cite{walborn2010spatial}
\begin{equation} \label{prob-pos}
P(\boldsymbol{\rho}_{s},\boldsymbol{\rho}_{I}) \propto \left|\int d \textbf{q}_{s} \ d \textbf{q}_{I}\  C(\textbf{q}_{s}, \textbf{q}_{I}) \ e^{i (\textbf{q}_{s}.\boldsymbol{\rho}_{s} + \textbf{q}_{I}.\boldsymbol{\rho}_{I})}\right|^{2}.
\end{equation}
The position correlation between the two photons is defined through this joint probability density. If $P(\boldsymbol{\rho}_{s},\boldsymbol{\rho}_{I})$ can be expressed as a product of a function of $\boldsymbol{\rho}_{s}$ and a function of $\boldsymbol{\rho}_{I}$, there is no position correlation. In the other extreme case, when the positions of the two photons are maximally correlated, the joint probability density is proportional to a delta function.
\par
The object is represented by it spatially dependent complex amplitude transmission coefficient, $T(\boldsymbol{\rho}_{o})$, where $\boldsymbol{\rho}_{o} \equiv M_{I} \boldsymbol{\rho}_{I}$ represents a point on an object and $M_I$ represents the magnification of the imaging system $B$. The single-photon counting rate (intensity) at a point, $\boldsymbol{\rho}_{c}$, on the camera is given by (c.f. \cite{viswanathan2021position}, Eqs. (10))
\begin{align} \label{photon count-final-expr}
\mathcal{R}(\boldsymbol{\rho}_{c}) \propto  \int d \boldsymbol{\rho}_{o} P\left(\frac{\boldsymbol{\rho}_{c}}{M_s}, \frac{\boldsymbol{\rho}_{o}}{M_I} \right) \left[1+ |T(\boldsymbol{\rho}_{o})|
\cos\left(\phi_{in}- \text{arg}\left\{T(\boldsymbol{\rho}_{o})\right\}\right)\right],
\end{align} 
where $\phi_{in}$ is the interferometer phase that can be varied experimentally, arg represents the argument of a complex number, and we have assumed for simplicity that the phases introduced by the imaging systems $A$, $B$, and $B'$ are not spatially dependent. The image of both absorptive and phase objects can be obtained from the interference pattern given by Eq. (\ref{photon count-final-expr}). It has been shown in Ref. \cite{viswanathan2021position} that the magnification of the imaging system is given by $M=M_s/M_I$.
\par
For a purely absorptive object, the amplitude transmission coefficient ($0\leq T\leq 1$) is a positive real number, i.e., $\text{arg}\left\{T(\boldsymbol{\rho}_{o})\right\}=0$. In this case, one needs to measure the photon counting rate (intensity) for the two cases $\cos(\phi_{in})=1$ and $\cos(\phi_{in})=-1$, which are respectively given by 
\begin{subequations}\label{photon-count-abs}
	\begin{align}
	&\mathcal{R}_{+}(\boldsymbol{\rho}_{c}) \propto  \int d \boldsymbol{\rho}_{o} P\left(\frac{\boldsymbol{\rho}_{c}}{M_s}, \frac{\boldsymbol{\rho}_{o}}{M_I} \right) \left[1+ |T(\boldsymbol{\rho}_{o})| \right], \label{photon-count-abs:a} \\
	&\mathcal{R}_{-}(\boldsymbol{\rho}_{c}) \propto  \int d \boldsymbol{\rho}_{o} P\left(\frac{\boldsymbol{\rho}_{c}}{M_s}, \frac{\boldsymbol{\rho}_{o}}{M_I} \right) \left[1- |T(\boldsymbol{\rho}_{o})| \right]. \label{photon-count-abs:b}
	\end{align} 
\end{subequations}
The image can be readily obtained by subtracting Eq. (\ref{photon-count-abs:b}) from Eq. (\ref{photon-count-abs:a}). The image is, therefore, given by
\begin{align} \label{image-abs obj}
G(\boldsymbol{\rho}_{c})=\mathcal{R}_{+}(\boldsymbol{\rho}_{c})-\mathcal{R}_{-}(\boldsymbol{\rho}_{c})\propto \int d\boldsymbol{\rho}_{o}
P\left(\frac{\boldsymbol{\rho}_{c}}{M_{s}}, \frac{\boldsymbol{\rho}_{o}}{M_{I}} \right) |T(\boldsymbol{\rho}_{o})|.
\end{align}
We call $G(\boldsymbol{\rho}_{c})$ the \emph{image function} which will be used to study the resolution.
\par
Alternatively, the image can also be obtained from the visibility of the interference pattern as shown in Ref. \cite{viswanathan2021position}. Here we use the intensity subtraction method because it simplifies the analysis. Furthermore, from the experimental perspective, this method is expected to be less affected by noise and intensity fluctuations because $\mathcal{R}_{+}(\boldsymbol{\rho}_{c})$ and $\mathcal{R}_{-}(\boldsymbol{\rho}_{c})$ can be measured simultaneously at the two outputs of the interferometer by the same camera. We stress that our results remain the same even if one obtains the image from the visibility.
\par 
The main features of the resolution can be captured by considering a purely absorptive object and, therefore, we restrict our analysis to this case only. In the following sections, we analyze the resolution of position correlation enabled QIUP in detail. 

\section{Analysis of resolution}\label{sec:reso-method}

\subsection{General Method}\label{subsec:genapp}
Equation (\ref{image-abs obj}) shows that the joint probability density ($P$), which governs the position correlation between the signal and the idler photons, appears in the photon counting rate (intensity) measured at the camera. This fact strongly suggests that the resolution will be limited by the position correlation between the two photons. In order the evaluate the integral in Eq. (\ref{image-abs obj}), one needs to assume a form for the joint probability density function $P$. We choose the form that pertains to a quantum state generated by non-degenerate spontaneous parametric down-conversion (SPDC) in a nonlinear crystal. The method to derive the form of this probability density function for degenerate SPDC is available in the literature (see, for example, \cite{monken1998transfer,Walborn2003multHOM,Tasca2009prop,grice2011spatial}). Thus, the probability density function for non-degenerate SPDC can be obtained by a straightforward generalization of this method and is found to be given by (see Appendix)
\begin{align}\label{Prob-Gauss-complete}
P (\boldsymbol{\rho}_{s},\boldsymbol{\rho}_{I}) &= A \exp \left[-\frac{2 }{w_{p}^{2}(\lambda_{I} + \lambda_{s})^{2}} \left|\lambda_{I}\boldsymbol{\rho}_{s} + \lambda_{s}\boldsymbol{\rho}_{I} \right|^{2} \right] \exp \left[-\frac{4 \pi}{L(\lambda_{I} + \lambda_{s})} \left|\boldsymbol{\rho}_{s}-\boldsymbol{\rho}_{I} \right|^{2} \right], 
\end{align}
where $\boldsymbol{\rho}_{I} = \boldsymbol{\rho}_{o}/M_{I}$, $\boldsymbol{\rho}_{s} = \boldsymbol{\rho}_{c}/M_{s}$, $w_p$ is the waist of the pump beam (assumed to have a Gaussian profile), $L$ is the crystal length, $\lambda_{s}$ and $\lambda_{I}$ are the wavelengths of signal and idler photons respectively, and $A$ is the normalization constant whose value is not relevant for our analysis (explicit form of $A$ is given in the Appendix). Typical experimental parameters are such that usually the first exponential term on the right-hand side of Eq. (\ref{Prob-Gauss-complete}) varies more slowly than the second term, and therefore the first term can be approximated by a constant, so that  Eq. (\ref{Prob-Gauss-complete}) can be reduced to the approximated form:
\begin{align} \label{Prob Gaussian}
P \left( \frac{\boldsymbol{\rho}_{c}}{M_{s}}, \frac{\boldsymbol{\rho}_{o}}{M_{I}}\right) \approx B \exp \left[-\frac{4 \pi}{L(\lambda_{I} + \lambda_{s})} \left|\frac{\boldsymbol{\rho}_{c}}{M_{s}} - \frac{\boldsymbol{\rho}_{o}}{M_{I}} \right|^{2} \right],
\end{align}
where we have applied the relations $\boldsymbol{\rho}_{I} = \boldsymbol{\rho}_{o}/M_{I}$ and $\boldsymbol{\rho}_{s} = \boldsymbol{\rho}_{c}/M_{s}$ and $B$ is a normalization constant. 
\par
In the following sections, we apply Eqs. (\ref{image-abs obj}) and (\ref{Prob Gaussian}) to quantitatively study the spatial resolution. In particular, we discuss the point spread function and the minimum resolvable distance. 

\subsection{Point spread function}\label{subsec:PSF}
The point spread function (PSF) is an important tool to understand the resolution of an imaging system. The PSF is the response of an imaging system to a point object \cite{goodman2005introduction}. It can be intuitively understood as the image of a point object for practical purposes. In the case of conventional imaging systems, the image can be fully characterized in terms of the complex optical field that interacts with the object. Although this is not the case for QIUP, Eq. (\ref{Prob Gaussian}) shows that the image is given by a linear transformation applied on the amplitude transmission coefficient of the object. Therefore, the concept of PSF can be readily applied to QIUP. 
\par
In order to determine the PSF, we consider a point object located at the point $(0,0)$ on the object plane, represented by 
\begin{align} \label{point obj}
T(\boldsymbol{\rho}_{o}) \propto \delta^{(2)}(\boldsymbol{\rho}_{o}) \equiv \delta(x_{o}) \delta(y_{o}),
\end{align}
where $x_{o}$ and $y_{o}$ represent the position along two mutually orthogonal Cartesian coordinate axes $X_o$ and $Y_o$, respectively, on the object plane.
Substituting from Eqs. (\ref{Prob Gaussian}) and (\ref{point obj}) into Eq. (\ref{image-abs obj}), we find that the corresponding image is given by the image function
\begin{align} \label{Vis-Point Obj}
G(\boldsymbol{\rho}_{c}) \propto \text{exp} \left[-\frac{4 \pi |\boldsymbol{\rho}_{c}|^2}{M_{s}^2L(\lambda_{I} + \lambda_{s})} \right]. 
\end{align}
We represent the PSF by the normalized form of this image function, i.e., by
\begin{align} \label{PSF}
\text{PSF}({\rho}_{c}) = \text{exp} \left[-\frac{4 \pi |\boldsymbol{\rho}_{c}|^2}{M_{s}^2L(\lambda_{I} + \lambda_{s})} \right]. 
\end{align}
\par
\begin{figure}[htbp]
	\centering 
	\includegraphics[width=0.9\linewidth]{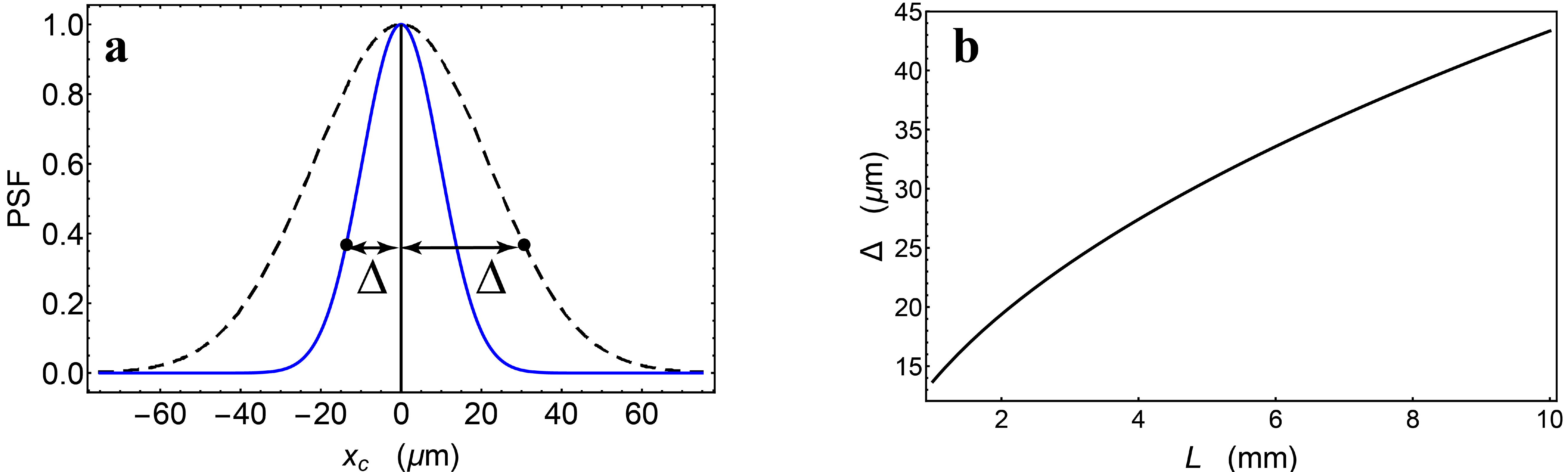}
	\caption{The point spread function (PSF). \textbf{a}, The PSF for a point object (pinhole) at the origin $(0,0)$ on the object plane is plotted against one of the camera coordinates ($x_{c}$) for two different crystal lengths, $L = 5$ mm (dashed curve) and $L = 1$ mm (solid curve).  \textbf{b}, The PSF spread ($\Delta$) is plotted against the crystal length ($L$). Shorter crystal length results in a stronger position correlation, which leads to a smaller PSF spread ($\Delta$). For both graphs we used the following parameters: $\lambda_{s} = 810$ nm, $\lambda_{I} = 1550$ nm.}
	\label{fig:PSFPlot}
\end{figure}

We define the spread $(\Delta)$ of the PSF by the distance at which the PSF drops to $1/e$ of its maximum value (Fig. \ref{fig:PSFPlot}a). We find from Eq. (\ref{PSF}) that 
\begin{align} \label{PSF spread}
\Delta = \frac{M_s}{2 \sqrt{\pi}}\sqrt{L (\lambda_{s} + \lambda_{I})}. 
\end{align}
We note that the crystal length ($L$) is in the PSF. Since the crystal length determines the position-correlation [see Eq. (\ref{Prob Gaussian})], the resolution is limited by the position correlation of the twin photons. A larger value of $L$ gives a weaker position correlation and we find from Eq. (\ref{PSF spread}) that in this case, the spread increases implying a reduced resolution.
\par
In Fig. \ref{fig:PSFPlot}a, we plot the PSF for two different values of the crystal length ($L$). The figure illustrates that the spread ($\Delta$) has a larger value for a larger crystal length. Figure \ref{fig:PSFPlot}b shows the dependence of $\Delta$ on $L$. This figure clearly illustrates that the spread of the PSF increases as $L$ increases, i.e., the resolution reduces as the position correlation between the twin photons become weaker.
\par
The spread ($\Delta$) of the PSF given by Eq. (\ref{PSF spread}) is given in terms of the object coordinates. In order to connect it to the minimum resolvable distance, we need to divide it by the magnification ($M$). It follows from (\ref{PSF spread}) and the formula for magnification ($M=M_s/M_I$) that
\begin{align} \label{PSF-spr-obj}
\frac{\Delta}{M} = \frac{M_I}{2 \sqrt{\pi}}\sqrt{L (\lambda_{s} + \lambda_{I})}, 
\end{align}
where $M_I$ is the magnification of the imaging system $B$ (Fig. \ref{fig:img-schm}). 
\par
Equation (\ref{PSF-spr-obj}) suggests that the resolution increases as the value of $M_I$ decreases. In Sec. \ref{subsec:mindist}, we verify this result by analyzing the minimum resolvable distance on the object plane.
\par
Equations (\ref{PSF spread}) and (\ref{PSF-spr-obj}) also show that the wavelengths of both the photons play a symmetric role in determining the resolution. This fact marks a striking difference with the momentum-correlation enabled QIUP for which the wavelength of the undetected photon alone characterizes the resolution \cite{fuenzalida2020resolution}. We show in Sec. \ref{subsec:mindist} that the same result is obtained from the analysis of the minimum resolvable distance. The wavelength dependence of resolution is discussed in further details in Sec. \ref{subsec:reso-wvln}.

\subsection{Minimum resolvable distance}\label{subsec:mindist}
We now analyze the minimum distance that the position correlation enabled QIUP can resolve. We consider two points, separated by a distance $d$, located on the object plane. Without any loss of generality we choose two radially opposite points located on axis $X_o$. The two points can therefore be represented by the amplitude transmission coefficient
\begin{align} \label{two radial points}
T(\boldsymbol{\rho}_{o})\equiv T(x_o,y_o) \propto \delta(y_{o})[\delta(x_{o}-d/2) + \delta(x_{o} + d/2)], 
\end{align}
where $x_{o}$ and $y_{o}$ represent the position along two mutually orthogonal Cartesian coordinate axes $X_o$ and $Y_o$, respectively, on the object plane.
\begin{figure}[htbp]
	\centering 
	\includegraphics[width=0.86\linewidth]{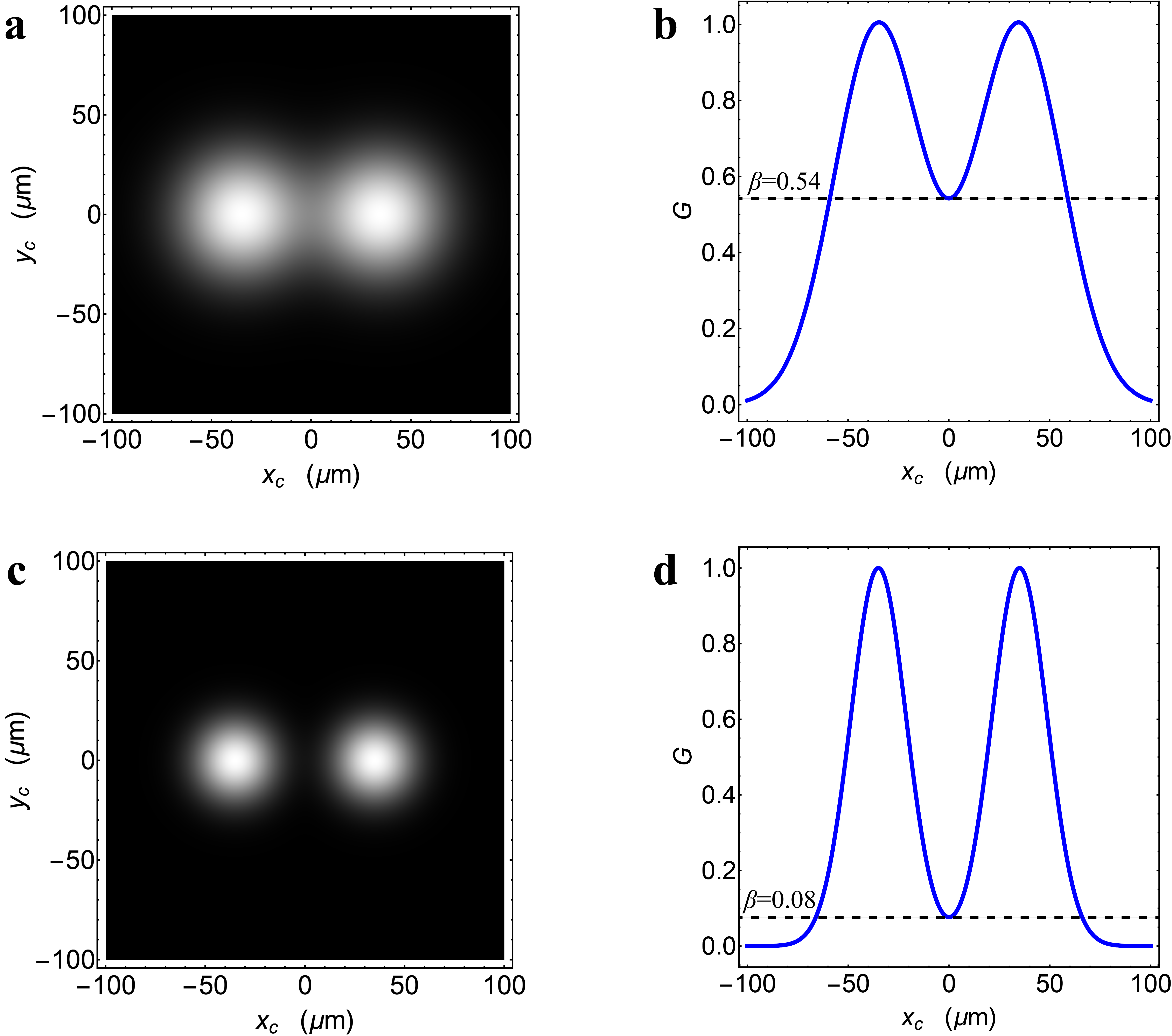}
	\caption{Enhancement of resolution with a stronger position correlation between twin photons. Two points separated by a distance of $d=70$ $\mu$m are imaged for weaker (top row) and stronger (bottom row) position correlation. \textbf{a}, Simulated camera image for crystal length $L=5$ mm. \textbf{b}, The image function, $G(x_{c},0)$, plotted against $x_c$ for $L=5$ mm. The ratio of its value at the dip to that at one of the peaks is $\beta\approx 0.54$. \textbf{c}, Simulated camera image of the same pair of points for crystal length $L=2$ mm shows higher resolution than in figure a. \textbf{d}, $G(x_{c},0)$ plotted against $x_c$ for $L=2$ mm. The ratio $\beta\approx 0.08$ quantitatively shows that the resolution in this case is higher than in figure b. (Choice of parameters: $\lambda_{s} = 810$ nm, $\lambda_{I} = 1550$ nm, and $M_{s} = M_{I} = 1$.)}
	\label{fig:VisbPlot}
\end{figure}
\par
It follows from Eqs. (\ref{image-abs obj}), (\ref{Prob Gaussian}), and (\ref{two radial points}) that the image of the two points is given by the image function
\begin{align} \label{Res radial points}
G(\boldsymbol{\rho}_{c})& \propto \exp\left[-\frac{4 \pi y_c^2}{M_s^2L(\lambda_{I} + \lambda_{s})} \right] \nonumber \\ &\times \left\{\exp\left[-\frac{4 \pi}{L(\lambda_{I} + \lambda_{s})} \left(\frac{x_{c}}{M_{s}} -\frac{d}{2 M_{I}} \right)^{2} \right] + \exp\left[-\frac{4 \pi}{L(\lambda_{I} + \lambda_{s})} \left(\frac{x_{c}}{M_{s}} +\frac{d}{2 M_{I}} \right)^{2} \right]\right\}.
\end{align}
Figures \ref{fig:VisbPlot}a and \ref{fig:VisbPlot}c illustrate the image of the same pair of points for weaker ($L=5$ mm) and stronger ($L=2$ mm) position correlations, respectively. Clearly, a stronger correlation results in a higher resolution. 
\par
In order to determine the minimum resolvable distance, it is enough to consider the values of $G(\boldsymbol{\rho}_{c})$ only along axis $X_c$ (i.e., $y_c=0$). Since the proportionality constant pertaining to Eq. (\ref{Res radial points}) is irrelevant for the analysis, we can set it equal to 1 and write
\begin{align} \label{Res-radial-points-2}
G(x_{c},0) = \exp\left[-\frac{4 \pi}{L(\lambda_{I} + \lambda_{s})} \left(\frac{x_{c}}{M_{s}} -\frac{d}{2 M_{I}} \right)^{2} \right] + \exp\left[-\frac{4 \pi}{L(\lambda_{I} + \lambda_{s})}  \left(\frac{x_{c}}{M_{s}} +\frac{d}{2 M_{I}} \right)^{2} \right].
\end{align}
If we plot $G(x_{c},0)$ against $x_c$, we get a double-humped curve (Figs. \ref{fig:VisbPlot}b and \ref{fig:VisbPlot}d). A measure of how well two points are resolved can be given by the ratio ($\beta$) of the value of $G$ at the dip ($G_{\text{dip}}$) to that at one of the peaks ($G_{\text{peak}}$), i.e., 
\begin{align} \label{ratio}
\beta\equiv \frac{G_\text{dip}}{G_\text{peak}}.
\end{align}
The lower the value of $\beta$, the better resolved the two points are, as can be seen by comparing Figs. \ref{fig:VisbPlot}b and \ref{fig:VisbPlot}d. 
\par
\begin{figure}[htbp]
	\centering 
	\includegraphics[width=0.9\linewidth]{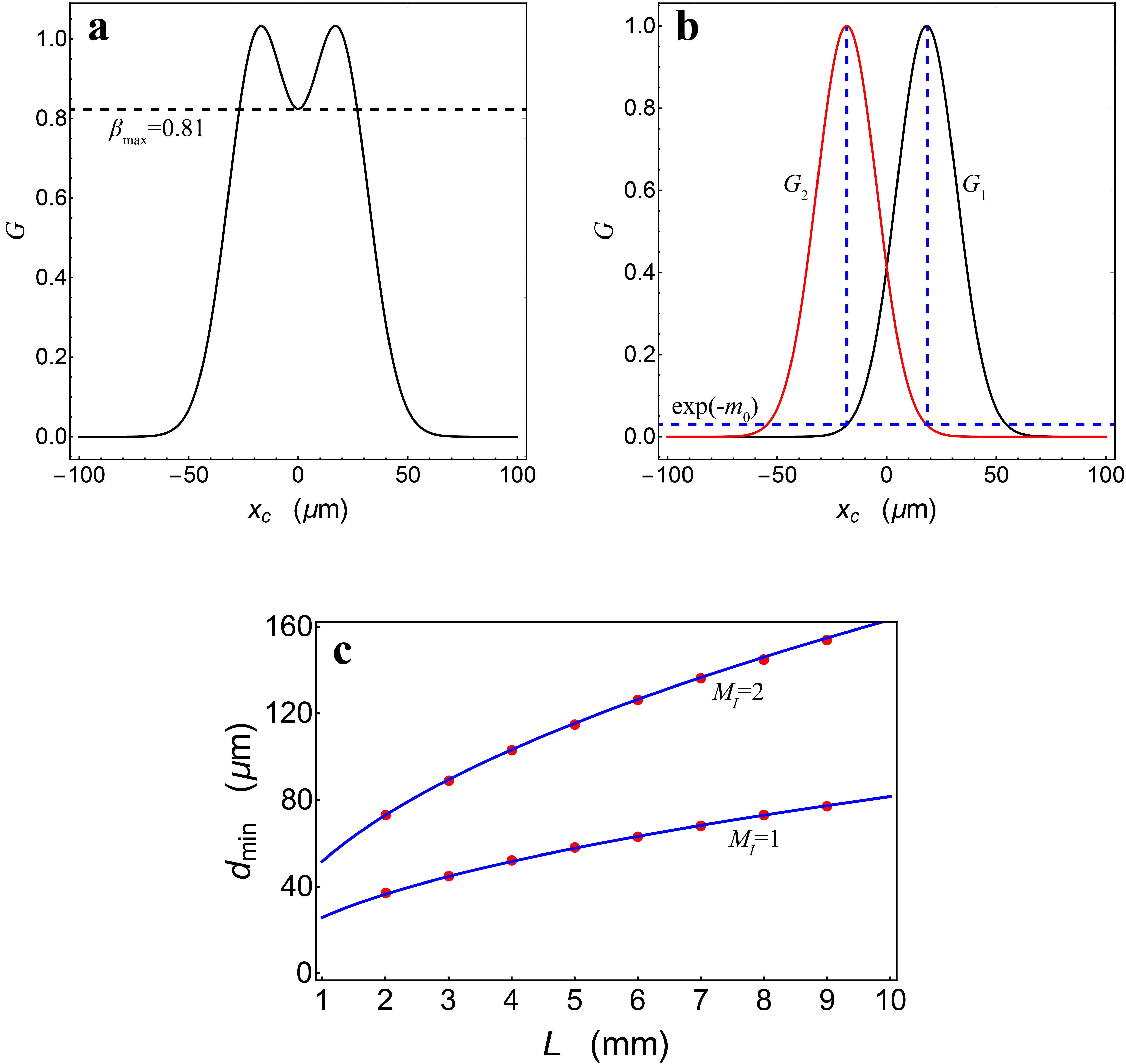}
	\caption{Minimum resolvable distance. \textbf{a}, The image function, $G(x_{c},0)$, for a pair of points that are just resolved. The ratio, $\beta$, attains the maximum allowed value $\beta_\text{max}=0.81$. The points are separated by $58$ $\mu$m on the object plane. (Choice of parameters: $\lambda_{s} = 810$ nm, $\lambda_{I} = 1550$ nm, $L=2$ mm, and $M_{I} = 1$.) \textbf{b}, Image functions, $G_1$ and $G_2$, for the same case considered in a. The value of $G_1$ is $\exp(-m_0)\approx 0.029$ at the point where $G_2$ attains its maximum value 1 and vice versa. \textbf{c}, The minimum resolvable distance ($d_\text{min}$) plotted against crystal length ($L$) for $M_I=1$ and $M_I=2$ using Eq. (\ref{mindist}) (solid lines). The filled circles represent simulated data points for a pair of square pinholes with side length $1$ $\mu$m. The minimum resolvable distance increases (i.e., resolution reduces) as the position correlation becomes weaker. The resolution also decreases as the imaging magnification, $M_I$, from the source to the object increases. (Remaining parameters are same as in a and b.)}\label{fig:min-resv-dist}
\end{figure}

The two points can no longer be resolved when $\beta$ exceeds a certain value, say, $\beta_\text{max}$. The two points are just resolved when $\beta=\beta_\text{max}$; in this case, the separation between the two points becomes the minimum resolvable distance (i.e., $d=d_\text{min}$). There is no strict rule to choose the value of $\beta_\text{max}$. For the purpose of illustration, we choose $\beta_\text{max}=0.81$. This value appears in the study of fine structure of the spectral lines with a Fabry-Perot interferometer (\cite{BW}, Sec. 7.6.3). In Fig. \ref{fig:min-resv-dist}a, we illustrate the image function, $G(x_{c},0)$, for a pair of points when they are just resolved. 
\par
We now note that the right-hand side of Eq. (\ref{Res-radial-points-2}) is a sum of two Gaussian terms. The first term corresponds to the image of the point $(d/2,0)$ and is denoted by $G_1(x_{c})$. Likewise, the second term that corresponds to the image of the point $(-d/2,0)$ is denoted by $G_2(x_{c})$. Each of these terms effectively represents the PSF discussed in Sec. \ref{subsec:PSF}. Clearly, $G_1(x_{c})$ and $G_2(x_{c})$ attain their maximum values at $x_c=Md/2$ and $x_c=-Md/2$, respectively. 
\par
Figure \ref{fig:min-resv-dist}b shows that, when $d=d_\text{min}$, the value of $G_1$ is $\exp(-m_0)$ at the point where $G_2$ attains its maximum value and vice versa. (The maximum value of $G_1$ and $G_2$ are normalized to unity. This implies the following relation: 
\begin{align}\label{resolv-cond}
G_1(-Md_\text{min}/2)=G_2(Md_\text{min}/2)=\exp(-m_0),
\end{align}
where $M=M_s/M_I$ and $m_0>0$. It is important to note that there is an one-to-one correspondence between $m_0$ and $\beta_\text{max}$. For a given value of $\beta_\text{max}$, the value of $m_0$ can be determined numerically. For $\beta_\text{max}=0.81$, we find that $m_0\approx 3.545$ and, consequently, $\exp(-m_0)\approx 0.029$. Figure \ref{fig:min-resv-dist}b illustrates the image functions of the two points for this case. 
\par
It immediately follows from Eqs. (\ref{Res-radial-points-2}) and (\ref{resolv-cond}) that 
\begin{align} \label{mindist}
d_{\text{min}} = n M_I \sqrt{L (\lambda_{I} + \lambda_{s})},
\end{align}
where $n=\sqrt{m_0/(4\pi)}$. Due to the one-to-one correspondence between $m_0$ and $\beta_\text{max}$, the value of $n$ can be numerically determined if the value of $\beta_\text{max}$ is given. For $\beta_\text{max}=0.81$, we find that $n\approx 0.53$.
\par 
Equation (\ref{mindist}) provides a quantitative measure of the minimum resolvable distance in the position correlation enabled QIUP. It shows that the minimum resolvable distance ($d_\text{min}$) is linearly proportional to the square root of the crystal length. Since a shorter crystal length implies a stronger position correlation between the twin photons, it becomes evident that a stronger position correlation between the twin photons results in a higher spatial resolution. Furthermore, the minimum resolvable distance is also linearly proportional to the magnification ($M_I$) of the imaging system, $B$, placed on the path of the undetected photon (see Fig. \ref{fig:img-schm}). Therefore, if the cross-section of the undetected beam (at source) is demagnified while illuminating the object, the spatial resolution enhances. We reached the same conclusions from Eq. (\ref{PSF-spr-obj}) while discussing the PSF.
\par
We use numerical simulations to illustrate these results further. We consider a pair of identical square apertures, each with side length $1$ $\mu$m, placed radially opposite on the $X_o$ axis (object plane). We choose nine values of the crystal length ($L$) and for each crystal length (i.e., fixed amount of position correlation), we choose two values of $M_I$. In each case, we numerically simulate the distance between the centers of the apertures by setting $\beta_\text{max}=0.81$. In Fig. \ref{fig:min-resv-dist}c, we compare these numerically simulated distances (data points represented by filled circles) with theoretically predicted minimum resolvable distances (solid curves) that are predicted by Eq. (\ref{mindist}). This figure clearly shows that the spatial resolution of position correlation enabled QIUP enhances (reduces) as the position correlation between the twin photons increases (decreases). The figure also illustrates that the resolution increases (reduces) if the cross-section of the undetected beam is demagnified (magnified) while illuminating the object.
\par
Equation (\ref{mindist}) also shows that the resolution is characterized by the wavelengths of both detected and undetected photons. We discuss the wavelength dependence of the resolution in the next section.

\subsection{Dependence of resolution on wavelength}\label{subsec:reso-wvln}
Our analysis thus far shows that when the resolution limit is dominated by the position correlation between the twin photons, the minimum resolvable distance is proportional to the square root of the sum of the wavelengths of the detected and undetected photons, i.e., $d_{\text{min}} \propto \sqrt{L(\lambda_{I}+\lambda_{s})}$. We note that both the detected wavelength ($\lambda_s$) and the undetected wavelength ($\lambda_I$) contribute symmetrically. Therefore, if one interchanges the two wavelengths, the resolution does not change. It also follows from Eqs. (\ref{PSF-spr-obj}) and (\ref{mindist}) that if one of the wavelengths is much longer than the other (e.g., optical wavelength and x-ray wavelength as demonstrated in \cite{PDCX-rays2017}), the resolution will in practice be limited by the longer wavelength.
\par
The presence of the square root in Eqs. (\ref{PSF-spr-obj}) and (\ref{mindist}) shows that very high resolution can be obtained if the wavelengths are chosen judiciously. For example, the resolution anticipated in state-of-art mid-infrared imaging systems ($1$ to $10$ $\mu$m) can be readily achieved if one performs an imaging experiment with the following standard experimental parameters: crystal length $L=1$ mm, detected wavelength $\lambda_s=647$ nm, undetected wavelength $\lambda_I=3$ $\mu$m, and $M_I=1/5$. In this case, the spread of the PSF (see Eq. (\ref{PSF-spr-obj})) is given by $\Delta/M\approx 3.4$ $\mu$m. If one determines the minimum resolving distance by setting $\beta_\text{max}=0.81$, one immediately finds from Eq. (\ref{mindist}) that $d_{\text{min}}\approx 6.4$ $\mu$m. Since in this case, the object is illuminated by the light of wavelength $\lambda=3$ $\mu$m, the resolution is of the order of $2\lambda$. 

\section{Comparison and Discussion}\label{sec:discuss}
\par
We now note the differences between our results and the resolution of momentum correlation enabled QIUP. The following central properties of the resolution of momentum correlation enabled QIUP are demonstrated in Ref \cite{fuenzalida2020resolution}: 1) the resolution enhances as the momentum correlation between the twin photons becomes stronger and is linearly proportional to the pump waist at the crystals; 2) the resolution is linearly proportional to the wavelength of the undetected photon that illuminates the object. In contrast, for position correlation enabled QIUP, we have shown that: 1) the resolution enhances as the position correlation between the twin photons becomes stronger and is inversely proportional to the square root of the length of the crystal; 2) the resolution is linearly proportional to the square root of the sum of the wavelengths of both detected and undetected photons. 
\par
We now compare the minimum resolvable distances obtained for position correlation enabled (near field) and momentum correlation enabled (far field) QIUP. As shown above, the former is given by $d_{\text{min}} \approx 0.53 M_I \sqrt{L (\lambda_{I} + \lambda_{s})}$. The minimum resolvable distance for the momentum correlation enabled QIUP is given by $d_{\text{min}} \approx 0.42 f_I\lambda_I/w_p$; this result is obtained by applying the theory presented in Ref. \cite{fuenzalida2020resolution} and following the procedure shown in Sec. \ref{subsec:mindist}. In this case, the presence of pump waist ($w_p$) shows that the resolution is limited by the momentum correlation between the twin photons. Furthermore, the presence of $\lambda_I$ shows that the resolution is characterized by the wavelength illuminating the object. The resolution, in the momentum correlation enabled case, can also be controlled by the focal length ($f_I$) of the positive lens that optically places the object in the far field relative to the sources. From the formulas for minimum resolvable distances, it becomes evident that one of the methods cannot, in general, be stated superior than the other. Their usefulness depends on the specific problem on hand. For example, the numerical values presented in Sec. \ref{subsec:reso-wvln} suggest that for applications to microscopy, the position correlation enabled QIUP may be advantageous over the momentum correlation enabled QIUP with the state-of-art technologies \footnote{In this context, see the recent microscopy experiments based on momentum correlation enabled \cite{kviatkovsky2020microscopy,paterova2020hyperspectral} and position correlation enabled \cite{kviatkovsky2021mid} QIUP.}.
\par
Finally, we touch upon the similarities between the resolution of QIUP and that of quantum ghost imaging (QGI). Like QIUP, the resolution of QGI is limited by the spatial correlation between the twin photons produced by the down-conversion source \cite{moreau2019imaging,fuenzalida2020resolution}. In QGI, when the object is placed in the far field of the source, the spatial resolution depends primarily on the wavelength which illuminates the object, whereas in the case where the object is placed at the near field of the source, both wavelengths play a role in defining the resolution\cite{chan2009two, rubin2008resolution, aspden2013epr, Aspden:15,Tasca:13}. It is important to note that unlike conventional imaging systems, the resolution limit imposed by spatial correlations in both QIUP and QGI cannot be improved by conventional optical techniques.

\section{Conclusion}\label{sec:concl}
We have studied the resolution of position correlation enabled quantum imaging with undetected photons (QIUP) in detail and have compared it with the case of momentum correlation enabled QIUP. We have considered the scenario in which the position correlation between the twin photons dominates the resolution limit. We have proved that the resolution enhances with position correlation between the twin photons and is linearly proportional to the square root of the sum of the wavelengths of the detected and undetected photons. We have further shown that the resolution enhances if one demagnifies the cross-section of the undetected beam while illuminating the object. Although we characterized the object by the transmission coefficient, the same method applies if one alternately characterized the object by the reflection coefficient. Therefore, our results also apply to a reflective object. 
\par
Our method and results provide a deeper understanding into the resolution limits of quantum imaging with undetected photons. We thus believe that our results will inspire further experiments and contribute to the field of quantum imaging as a whole. 

\section*{Acknowledgement.} \noindent B.V. and M.L. acknowledge support from College of Arts and Sciences and the Office of the Vice President of Research, Oklahoma State University. G.B.L. acknowledges support from the Brazilian National Council for Scientific and Technological Development (CNPq) and from the Coordenação de
Aperfeiçoamento de Pessoal de Nível Superior (CAPES - Brazil) -- Finance Code 001. 

\section*{Appendix}
\setcounter{equation}{0}
\renewcommand{\theequation}{A-\arabic{equation}}
In this appendix, we derive the form of the joint probability density function, $P(\boldsymbol{\rho}_{s}, \boldsymbol{\rho}_{I})$, used in Eq. (\ref{Prob-Gauss-complete}). The joint probability density $P(\boldsymbol{\rho}_{s}, \boldsymbol{\rho}_{I})$ for non-degenerate SPDC in a nonlinear crystal can be obtained by generalizing the existing method for degenerate SPDC \cite{monken1998transfer,Walborn2003multHOM,Tasca2009prop,grice2011spatial}. 
\par
The  coefficient $C(\textbf{q}_{s},\textbf{q}_{I})$ in Eq. (\ref{state-spdc}), can be expressed in the following form (cf. \cite{hochrainer2017quantifying}, Supplementary Information, Eqs. $(S2)$ and $(S3)$):
\begin{align} \label{C-fn gen form}
C(\textbf{q}_{s},\textbf{q}_{I}) &\propto \xi(\textbf{q}_{s} + \textbf{q}_{I}) \ \gamma \left(\textbf{q}_{s}, \textbf{q}_{I} \right),
\end{align}
where $\xi$ is the angular spectrum of the pump and the phase matching function, $\gamma$, has the form
\begin{align} \label{sinc fn}
\gamma(\textbf{q}_{s},\textbf{q}_{I}) &= \text{sinc}\left(\frac{L \lambda_{p} \lambda_{s}}{8 \pi \lambda_{I}} \left| \textbf{q}_{s} - \frac{\lambda_{I}}{\lambda_{s}} \textbf{q}_{I} \right|^{2} \right),
\end{align}
where $\text{sinc}(x) = \sin(x)/x$, $L$ is the length of the nonlinear crystal, $\lambda_{p}$, $\lambda_{s}$ and $\lambda_{I}$ are the wavelengths of the pump, signal and idler photons, respectively. 
\par
We assume that the angular spectrum of the pump has a Gaussian profile, i.e., 
\begin{align} \label{Gaussian pump}
\xi(\textbf{q}_{s}+ \textbf{q}_{I}) &= \exp\left(-\frac{1}{4}|\textbf{q}_{s} + \textbf{q}_{I}|^{2} w_{p}^{2} \right),
\end{align}
where $w_{p}$ is the waist of the pump beam. Following the standard practice for the degenerate SPDC (see for example, \cite{Law-Eberly,Tasca2009prop,schneeloch2016introduction}), we approximate the sinc form of $\gamma$ by a Gaussian function and write
\begin{align} \label{Gauss Phase Match}
\gamma(\textbf{q}_{s},\textbf{q}_{I}) &= \exp\left(- \frac{L\lambda_{p} \lambda_{s}}{8 \pi \lambda_{I}} \left|\textbf{q}_{s} - \frac{\lambda_{I}}{\lambda_{s}}\textbf{q}_{I}\right|^{2}\right). \end{align}
From Eqs. (\ref{C-fn gen form}), (\ref{Gaussian pump}), and (\ref{Gauss Phase Match}) we have
\begin{align} \label{C-fn Gaussian}
C(\textbf{q}_{s}, \textbf{q}_{I}) \propto \exp\left(-\frac{1}{4}|\textbf{q}_{s} + \textbf{q}_{I}|^{2} w_{p}^{2} \right) \exp\left(- \frac{L\lambda_{p} \lambda_{s}}{8 \pi \lambda_{I}} \left|\textbf{q}_{s} - \frac{\lambda_{I}}{\lambda_{s}}\textbf{q}_{I}\right|^{2}\right).
\end{align}
On substituting from Eq. (\ref{C-fn Gaussian}) into Eq. (\ref{prob-pos}) of the main text, using the relation $\lambda_{p} \approx \lambda_{s} \lambda_{I} / (\lambda_{s} + \lambda_{I})$, and normalizing the probability density function, we find that
\begin{align} \label{Prob Gaussian pump/phase}
P \left( \boldsymbol{\rho}_{s}, \boldsymbol{\rho}_{I}\right)&= \frac{8}{\pi L w_{p}^{2}(\lambda_{s} + \lambda_{I})}\ \exp \left[-\frac{4 \pi}{L(\lambda_{I} + \lambda_{s})} \left|\boldsymbol{\rho}_{s} - \boldsymbol{\rho}_{I} \right|^{2} \right] \cr
&\times \exp \left[-\frac{2 }{w_{p}^{2}(\lambda_{I} + \lambda_{s})^{2}} \left|\lambda_{I}\boldsymbol{\rho}_{s} + \lambda_{s}\boldsymbol{\rho}_{I} \right|^{2} \right].
\end{align}

\bibliography{references-reso}

\end{document}